\documentclass[A4paper,12pt,doublespace]{article}
\usepackage[a4paper,inner =
1.7cm,outer=2.7cm,top=2cm,bottom=2cm,bindingoffset=1.2cm]{geometry}
\usepackage{verbatim}
\usepackage{graphicx}
\usepackage{placeins}
\usepackage{authblk}
\begin{document}

\title{Identification of Biomarkers Driving Blood Cell Development}
\author[1,2,*]{Maryam Nazarieh}
\author[2,*]{Volkhard Helms}
\affil[1]{Graduate School of Computer Science, Saarland University, Saarbruecken, Germany}
\affil[2]{Center for Bioinformatics, Saarland University, Saarbruecken, Germany}
\affil[*]{volkhard.helms@bioinformatik.uni-saarland.de,maryam.nazarieh@bioinformatik.uni-saarland.de}
\maketitle
\abstract{
A blood cell lineage consists of several consecutive developmental stages
from the pluripotent or multipotent stem cell to a particular stage
 of terminally differentiated cells.
There is considerable interest in identifying the key regulatory genes that 
govern blood cell development from the gene expression data without
considering the underlying network between transcription factors (TFs) and their
target genes.
In this study, we introduce a novel expression pattern that key regulators
expose along the differentiation path. We deploy this pattern to identify the
cell-specific key regulators responsible for the development. 
As proof of concept, we consider this approach to data on six developmental
stages from mouse embryonic stem cells to terminally differentiated
macrophages.}
\section{Introduction}
\begin{figure}
	\includegraphics[width=\textwidth]{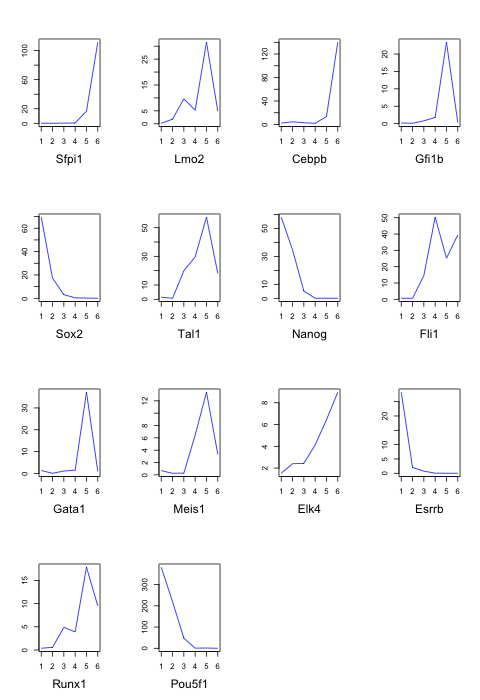}
 \caption{Expression of the global regulators driving hematopoietic
 specification for all six stages of blood development from ESCs.}
\label{fig:fig1}
\end{figure}

The problem of identification of key player genes in gene regulatory
networks (GRNs) mainly addressing cell cycle process has been explained in our
previous study \cite{Nazarieh2016}.
In that study, we considered the set of genes that form the minimum
dominating set (MDS) and minimum connected dominating set (MCDS) in the largest
connected component of the underlying network as the key player genes.
In this study, we focus on the identification of key regulatory genes
driving blood cell development based on the pattern of gene expression data that
they expose for particular stage along the cell development. 
We got inspired by the pattern of gene expression of the 
global regulators where the set of global regulators introduced in \cite{GOODE2016572}.
In that study, a dynamic core regulatory network is presented by multi-omics analysis where the network consists
 of 16 TFs that drive cellular differentiation from ESC to
 terminally differentiated macrophages. Embryonic stem cells (ESCs),
 Mesoderm (MES), Hemangioblast (HB), Hemogenic endothelium (HE),
 Hematopoietic progenitor (HP) and Macrophages (MAC) constitute the six
 consecutive developmental stages.
 These global regulators which have separate representations for all the
 developmental stages are connected to each other through a dense network. The network is inferred through a deep
 muli-omics analysis by generating data for RNA-seq, ChIP-seq and DNA-seq
 measuring global gene expression, chromatin accessibility, histone
 modifications and transcription factor binding data. 
 Although this network is preserved across all the developmental stages, but
 the global regulators have separate representations in each stage, e.g, Nanog, Esrrb, Pou5f1 and Sox2 are
 highly expressed in embryonic stem cell (ESC), where their expression level
 reach to the least level in the macrophages with gradual decrease along the path.
 Plotting the expression level of global regulators for all six stages of
 development with respect to the highest expressed regulators in particular
 cells inspired us to identify the cell-specific key regulators driving blood
 cell development from the gene expression data, see Figure \ref{fig:fig1}.
  The main idea is that a key player associated with a certain
differentiated cell has the highest expression value in that cell among all the
cells in the lineage.
Therefore, the key genes associated with an undifferentiated cell at the top of
the lineage have the highest expression level initially, however, the
expression level decreases gradually.
This pattern of expression looks different for the key players of the cells
in the middle stages.
In the middle stages, the expression value of a key player gene starts with
the least expression value at the top of the hierarchy with a gradual
increase along the differentiation path that reaches to the maximum value in the target
cell and then decreases gradually till it reaches to the terminally
differentiated cells. The genes with the highest expression values at the
terminally differentiated cells among all the cells with the gradual increase in
expression level from the initial stage are considered as candidates for the key
players in those cells.\\

\section{Results}
\begin{figure}
	\includegraphics[width=\textwidth]{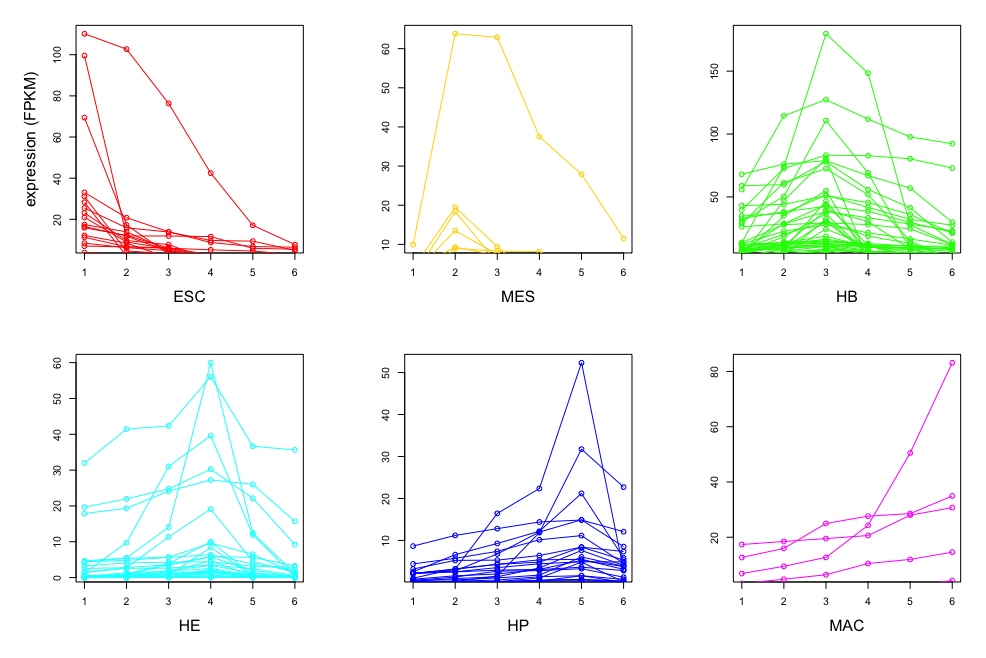}
 \caption{Depiction of the set of TFs for 6 stages of development
 following the pattern of expression.}
\label{fig:fig2}
\end{figure}

We applied our supervised method to data on six developmental stages from ESCs
to terminally differentiated macrophages compiled by (Goode et al. 2016)
\cite{GOODE2016572}.
The key players in the ESCs have the highest expression
level at the initial stage. The expression level decreases gradually when
it reaches to the terminally differentiated cells. For other cells along the
lineage, the related genes have the least expression value at the ESCs and the
highest in the target cells with a gradual decrease till it reaches to the final
stage, see Figure \ref{fig:fig2}.
This approach identified 332 genes including 22 TFs based on the
above-mentioned expression pattern in the ESC stage including Sox2 and Esrrb.
Pou5f1 has almost the similar pattern such that it has the highest expression value
 at the ESC stage, gradually decreases till HE, then less than
one unit increase in HP and again reduction till MAC (ESC: 382.02500,
MES: 219.5270000, HB: 47.8558000, HE: \textbf{0.687413}, HP: \textbf{1.19097},
MAC: 0.000000).
For Nanog the exception occurred at stage HE such that the expression value at
HE is slightly less than HP and MAC (ESC: 58.0737, MES: 34.6255, HB: 5.18978, HE: \textbf{0.055367}, HP:
\textbf{0.0734063}, MAC: 0.0159039).
Similarly, 18 TFs were identified for MES and 57, 29, 21 and 10 for HB, HE, HP
and MAC among 332, 702, 506, 201 and 309 genes. The set of TFs were extracted
from TF-gene interaction database \cite{TRRUST}. 

\subsection{Functional annotation}
In this work, the biological function of the genes in each stage 
was evaluated using the enrichment analysis tool provided at the DAVID portal of
NIH (version 6.8) based on the functional categories in GO Direct
\cite{Huang2009}.
p-values below the threshold 0.05 obtained by the hypergeometric test were
adjusted for multiple testing using the Benjamini \& Hochberg (BH) method
\cite{Benjamini1995}.

\subsection{Enrichment analysis fo the set of identified TFs}
Enrichment analysis for the set of identified TFs in the ESC stage yielded the
enriched biological process GO terms and KEGG pathways listed in Table S1. The list
includes GO terms such as GO:0048863, GO:0007275, GO:0007492 and GO:0030154
annotated to stem cell differentiation, multicellular organism development,
endoderm development and cell differentiation, respectively. Moreover, the set
of genes (Onecut1, Esrrb, Id1, Sox2, Zic3) are related to the KEGG pathway:
mmu04550 annotated to signaling pathways regulating pluripotency of stem cells.
Tables S2-S6 list the enriched GO terms and KEGG pathways for the identified
TFs in MES, HB, HE, HP and MAC. The lists include new specialized GO terms in
addition to some of the aforementioned terms such as GO:0001569, GO:0045165,
GO:0007507 and GO:0030097 annotated to patterning of blood vessels, cell fate
commitment, heart development and hemopoiesis, respectively and also a KEGG
pathway:
mmu05221 annotated to acute myeloid leukemia.

\subsection{Enrichment analysis for the set of identified target genes}
We inferred the set of target genes for the set of identified TFs from TF-gene
interaction database TRRUST \cite{TRRUST}.
Enrichment analysis for the set of identified target genes in the ESC stage
yielded the enriched biological process GO terms and KEGG pathways listed in Table S7. The list
includes GO terms such as GO:0042127, GO:0007275, GO:0048863, GO:0030154,
GO:0045165, GO:0048468 and GO:0008283 annotated to regulation of cell
proliferation, multicellular organism development, stem cell differentiation, 
cell differentiation, cell fate commitment, cell development and cell
proliferation, respectively.
Tables S8-S12 list the enriched GO terms for the target genes in other
developmental stages. In addition to common GO terms, distinct GO terms such as
GO:0061312 and GO:0016055 with annotations BMP signaling pathway involved in heart development and Wnt
signaling pathway are added in MES stage. More specialized GO terms appear in
later stages HB and HE and HP such as GO:0001889, GO:0002326, GO:0043583 and
GO:0001654 with annotations liver development, B cell lineage commitment, ear
development and eye development, respectively.

\subsection{Prioritization of the identified TFs}
\begin{figure}[h!]
	\includegraphics[width=\textwidth]{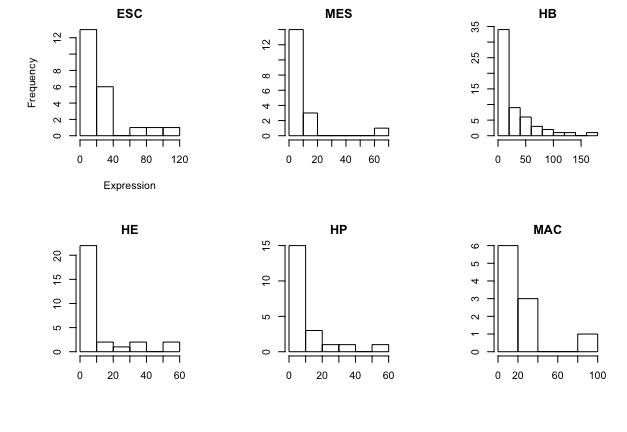}
 \caption{Histograms of cell-specific TFs in the blood cell lineage.}
\label{fig:fig3}
\end{figure}

\begin{figure}[h!]
	\includegraphics[width=\textwidth]{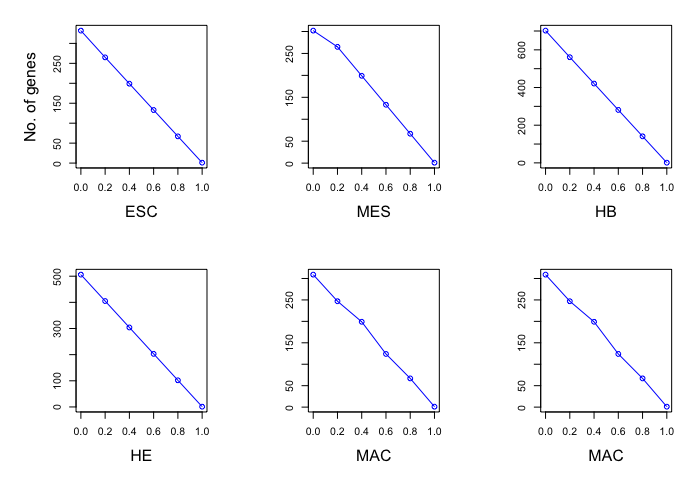}
 \caption{Number of stage-specific genes based on different quantiles.}
\label{fig:fig4}
\end{figure}

\begin{figure}[h!]
	\includegraphics[width=\textwidth]{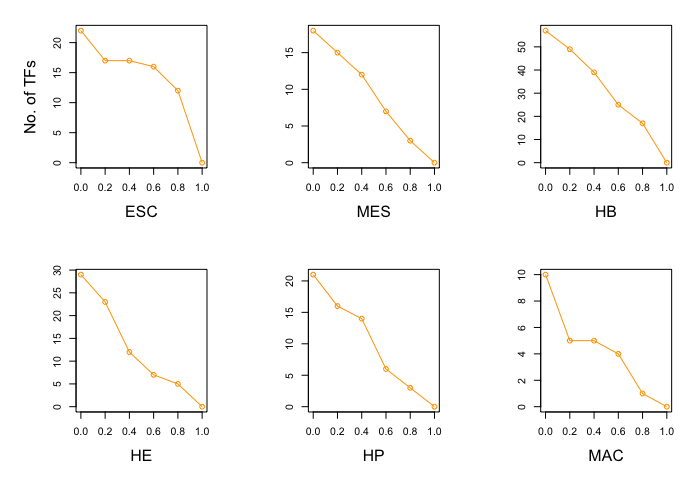}
 \caption{Number of stage-specific TFs based on different quantiles.}
\label{fig:fig5}
\end{figure}

Although the identified TFs in each particular developmental stage follow the
aforementioned expression pattern, but they expose different expression levels.
Histograms in Figure \ref{fig:fig3} show the frequency of TFs based on
expression level. As the expression level increases, the number of associated
TFs reduces.
\FloatBarrier
To prioritize the set of identified TFs, we considered a statistical measure
such as quantile. We used 5-quantiles that divides the set of genes into 5 equal
groups, see Figure \ref{fig:fig4}. For each cut point, the set of identified
genes were considered that are greater or equal the quantile value. Figure
\ref{fig:fig5} shows the number of TFs among the genes. The plots in figure
\ref{fig:fig5} show all the identified TFs initially with constant decrease
while increasing the quantile value.
\FloatBarrier
\FloatBarrier
\section{Conclusion}
We proposed an expression pattern for the identification of biomarkers driving
blood cell development. This expression pattern helps to infer the patient-specific
biomarkers by considering just the expression level across the cell lineage. No
prior knowledge of the regulatory interaction between genes is required to
derive the set of biomarkers. We have prioritized the set of identified
cell-specific TFs and genes based on the expression pattern.
Enrichment analysis suggests that these biomarkers can be potential targets 
for disease-related biomarkers such as leukaemia.
\bibliographystyle{unsrt}
\bibliography{KPE} 
\end{document}